\documentclass[submission,copyright,creativecommons]{eptcs}
\providecommand{\event}{PLACES 2020} 
\hypersetup{hidelinks}
\usepackage{wrapfig}
\usepackage{graphicx}
\usepackage{subcaption}
\usepackage{stmaryrd}
\usepackage[strings]{underscore}
\usepackage{color}
\usepackage{etoolbox}
\newtoggle{draft}
\definecolor{purple}{rgb}{0.65, 0.12, 0.82}
\definecolor{darkgray}{rgb}{.4,.4,.4}

\usepackage{listings}
\usepackage[noabbrev, capitalise]{cleveref}
\lstdefinelanguage{Scribble}{
  keywords={module, type, from, as, global,protocol, role, to, choice, at, or, do},
  keywordstyle=\color{blue}\bfseries,
  identifierstyle=\color{black},
  sensitive=false,
  comment=[l]{//},
  morecomment=[s]{/*}{*/},
  commentstyle=\color{darkgray}\ttfamily,
  stringstyle=\color{purple}\ttfamily,
  morestring=[b]',
  morestring=[b]",
}
\lstdefinelanguage{JavaScript}{
  keywords={typeof, new, true, false, catch, function, return, null, catch, switch, var, if, in, while, do, else, case, break, class, export, boolean, throw, implements, import, this, const, let, extends},
  keywordstyle=\color{blue}\bfseries,
  identifierstyle=\color{black},
  sensitive=false,
  comment=[l]{//},
  morecomment=[s]{/*}{*/},
  commentstyle=\color{darkgray}\ttfamily,
  stringstyle=\color{purple}\ttfamily,
  morestring=[b]',
  morestring=[b]"
}
\usepackage{caption}

\toggletrue{draft}

\title{Generating Interactive WebSocket Applications in TypeScript}
\author{Anson Miu
\institute{Imperial College London}
\and
Francisco Ferreira
\institute{Imperial College London}
\and
Nobuko Yoshida
\institute{Imperial College London}
\and
Fangyi Zhou
\institute{Imperial College London}
}
\def\titlerunning{Generating Interactive WebSocket Applications in TypeScript}
\def\authorrunning{A. Miu, F. Ferreira, N. Yoshida \& F. Zhou}
\begin{document}
\maketitle
\begin{abstract}
Advancements in mobile device computing power have made interactive web
applications possible, allowing the web browser to render contents dynamically
and support low-latency communication with the server.
This comes at a cost to the developer, who now needs to reason more about
correctness of communication patterns in their application as web applications
support more complex communication patterns.

Multiparty session types (MPST) provide a framework for verifying conformance
of implementations to their prescribed communication protocol.
Existing proposals for applying the MPST framework in application developments
either neglect the event-driven nature of web applications, or lack
compatibility with industry tools and practices, which discourages mainstream
adoption by web developers.

In this paper, we present an implementation of the MPST framework for
developing interactive web applications using familiar industry tools using
TypeScript and the \textit{React.js} framework.
The developer can use the Scribble protocol
language to specify the protocol and use the Scribble toolchain to validate
and obtain the \emph{local protocol} for each role.
The local protocol describes the interactions of the global communication
protocol observed by the role.
We encode the local protocol into TypeScript types, catering for server-side
and client-side targets separately.
We show that our encoding guarantees that only implementations which conform to
the protocol can type-check.
We demonstrate the effectiveness of our approach through a web-based
implementation of the classic \textit{Noughts and Crosses} game from an MPST
formalism of the game logic.

\end{abstract}

\section{Introduction}

Modern interactive web applications aim to provide a highly responsive user
experience by minimising the communication latency between clients and servers.
Whilst the HTTP request-response model is sufficient for retrieving static
assets, applying the same stateless communication approach for interactive use
cases (such as real-time multiplayer games) introduces undesirable performance
overhead.
Developers have since adopted other communication transport abstractions over
HTTP connections such as the WebSockets protocol \cite{WebSocketRFC} to enjoy
low-latency full-duplex client-server communication in their applications over
a single persistent connection.
Enabling more complex communication patterns caters for more interactive use
cases, but introduces additional correctness concerns to the developer.

Consider a classic turn-based board game of \textit{Noughts and Crosses}
between two players.
Both players, identified by either \emph{noughts (O's)} or \emph{crosses (X's)} respectively, take
turns to place a mark on an unoccupied cell of a 3-by-3 grid until one player
wins (when their markers form one straight line on the board) or a stalemate is
reached (when all cells are occupied and no one wins).
A web-based implementation may involve players connected to a game server via
WebSocket connections.
The players interact with the game from their web browser, which shows a
\textit{single-page application} (SPA) of the game client written in a popular
framework like \textit{React.js} \cite{React}.
SPAs feature a single HTML page and dynamically render content via JavaScript
in the browser.
Players take turns to make a move on the game board, which sends a message to
the server.
The server implements the game logic to progress the game
forward until a result (either a win/loss or draw) can be declared, where
either the move of the other player or the game result is sent to players.

Whilst WebSockets make this web-based implementation possible, they introduce
the developer to a new family of communication errors, even for this simple
game.
In addition to the usual testing for game logic correctness, the
developer needs to test against \textit{deadlocks} (e.g.\ both players waiting
for each other to make a move at the same time) and \textit{communication
 mismatches} (e.g.\ player 1 sending a boolean to the game server instead of
the board coordinates).
The complexity of these errors correlates to the
complexity of the required tests and scales with the complexity of
communication patterns involved.

\textit{Multiparty Session Types} (MPST) \cite{MPST} provide a framework for
formally specifying a structured communication pattern between concurrent
processes and verifying implementations for correctness with respect to the
communications aspect.
By specifying the client-server interactions of our game as an MPST protocol
and verifying the implementations against the protocol for conformance,
MPST theory guarantees well-formed implementations to be free from
communication errors.

We see the application of the MPST methodology to generating interactive
TypeScript web applications to be an interesting design space
--- to what extent can the MPST methodology be applied to
deliver a workflow where developers use the generated TypeScript APIs in
their application to guarantee protocol conformance by construction?
Such a workflow would ultimately decrease the overhead for incorporating MPST
into mainstream web development, which reduces development time by programmatically
verifying communication correctness of the implementation.

\paragraph{Contributions}
This paper presents a workflow for developing type-safe interactive SPAs
motivated by the MPST framework:
\textbf{(1)} An endpoint API code generation
workflow targeting TypeScript-based web applications for multiparty sessions;
\textbf{(2)} An encoding of session types in server-side TypeScript that
enforces static linearity;
and \textbf{(3)} An encoding of session types in
browser-side TypeScript using the React framework that guarantees affine usage
of communication channels.


\section{The Scribble Framework}
\label{section:scribble}

Development begins with specifying the expected communications between
participants as a \textit{global protocol} in Scribble \cite{Scribble}, a
MPST-based protocol specification language and code generation toolchain.
We specify the \textit{Noughts and Crosses} game as a Scribble protocol in
\cref{lst:game}.
In the protocol, the role \texttt{Svr} stands for the Server, and the roles
\texttt{P1} and \texttt{P2} stand for the two Players respectively.

\begin{figure}[!ht]
\begin{lstlisting}[
	language=Scribble
]
module NoughtsAndCrosses;
type <typescript> "Coordinate" from "./Types" as Point;	// Position on board

global protocol Game(role Svr, role P1, role P2) {
  Pos(Point) from P1 to Svr;
  choice at Svr {
    Lose(Point) from Svr to P2; Win(Point) from Svr to P1;
  } or {
    Draw(Point) from Svr to P2; Draw(Point) from Svr to P1;
  } or {
    Update(Point) from Svr to P2; Update(Point) from Svr to P1;
    do Game(Svr, P2, P1); // Continue the game with player roles swapped
  }
}
\end{lstlisting}
\captionof{lstlisting}{\textit{Noughts and Crosses} in a Scribble protocol.}
\label{lst:game}
\end{figure}

We leverage the Scribble toolchain to check for protocol
well-formedness.
This directly corresponds to multiparty session
type theory \cite{FeatherweightScribble}:
a Scribble protocol maps to some \textit{global type}, and the Scribble
toolchain implements the algorithmic projection defined in \cite{MPST} to
derive valid local type \textit{projections} for all participants.
We obtain a set of \textit{endpoint protocols} (corresponds to \emph{local
  types}) --- one for each role from a
well-formed global protocol.
An endpoint protocol only preserves the interactions defined by the global
protocol in which the target role is involved, and corresponds to an equivalent
\textit{Endpoint Finite State Machine} (EFSM) \cite{ICALP13CFSM}.
The EFSM holds information about the permitted IO actions for the role.
We use the EFSMs as a basis for API generation and adopt the formalisms in
\cite{Hybrid2016}.

\section{Encoding Session Types in TypeScript}

Developers can implement their application using APIs generated from the EFSM
to guarantee correctness by construction.
Our approach integrates the EFSM into the development workflow by encoding
session types as TypeScript types.
Communication over the WebSocket protocol introduces additional constraints:
communication is always initiated in the front-end and driven by user interactions,
whilst back-end roles can only accept connections.
This motivates our design of encoding the session types differently for server
(\cref{section:server}) and client (\cref{section:browser}) targets.

\subsection{Server-Side API Generation}
\label{section:server}


\begin{wrapfigure}{r}{0.6\textwidth}
  \vspace{-5mm}
  \begin{center}
    \includegraphics[width=0.58\textwidth]{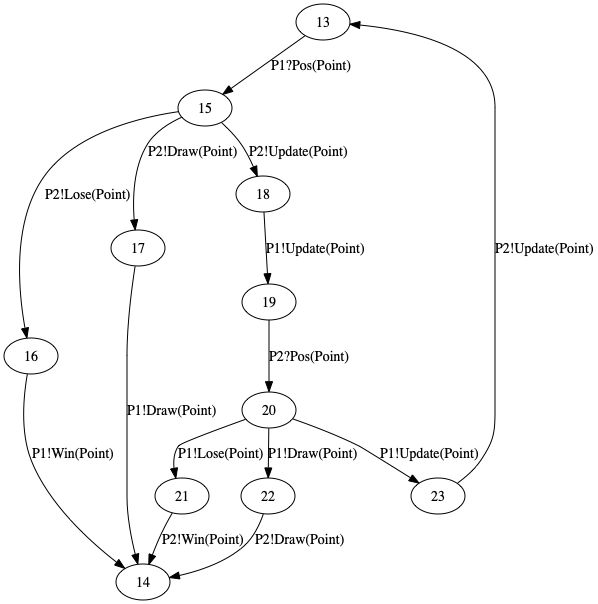}
  \end{center}

  \vspace{-5mm}
  \captionof{figure}{EFSM for \texttt{Svr}.}
  \label{fig:efsmsvr}
\vspace{-1cm}
\end{wrapfigure}

We refer to the \texttt{Svr} EFSM (\cref{fig:efsmsvr}) as a running example in
this section.
For server-side targets, we encode EFSM states into TypeScript types and
consider branching (receiving) and selection (sending) states separately.
We assign TypeScript encodings of states to their state identifiers in the
EFSM, providing syntactic sugar when referring to the successor state when
encoding the current state.
For any state $S$ in the EFSM, we refer to the TypeScript type alias of its
encoding as $\llbracket S \rrbracket$.
We outline the encoding below using examples from the
\textit{Noughts and Crosses} game (\cref{lst:svr}).

\paragraph{Branching State}
We consider a receiving state as a unary branching state for conciseness.
A branching state is encoded as an \textit{object literal}
\cite{TypeScriptSpec} (a record type), with each branch $i \in I$ ($I$ denoting set of all
branches), corresponding to a member field.
A branch expecting to receive a message labelled $\texttt{label}_i$ carrying
payload of type $\texttt{T}_i$ with successor state $S_i$ is encoded as an
\textit{member field} named $\texttt{label}_i$ of function type
$(payload:\texttt{T}_i) \to \llbracket S_i \rrbracket$.
The developer implements a branching operation by passing callbacks for each
branch, parameterised by the expected message payload type for that branch.

\paragraph{Selection State}
We consider a sending state as a unary selection state for conciseness.
A selection state is encoded as a \textit{union type}
\cite{TypeScriptSpec} of internal choice encodings: each internal choice $i \in
I$ ($I$ denoting set of all choices), sending a message labelled
$\texttt{label}_i$ carrying payload of type $\texttt{T}_i$ with successor state
$S_i$ is encoded as a \textit{tuple type} of \texttt{[Labels.label$_i$, T$_i$,
  $\llbracket S_i \rrbracket$]}.
The developer implements a selection operation by passing the selected label
and payload to send in the message.
We generate a \textit{string enum} (named \texttt{Labels}) wrapping the labels
in the protocol.

\begin{figure}[ht]
\begin{lstlisting}[language=JavaScript]
export type S13 = { Pos: (payload: Point) => S15 };
export type S15 = [ Labels.Lose, Point, S16 ]
                | [ Labels.Draw, Point, S17 ]
                | [ Labels.Update, Point, S18 ];
\end{lstlisting}
\captionof{lstlisting}{Example encodings from \textit{Noughts and Crosses} \texttt{Svr} EFSM.}
\label{lst:svr}
\end{figure}

In the case of \cref{lst:svr}, the developer is expected to implement
\texttt{S13} which handles the \texttt{Pos} message sent by \texttt{P1},
and the code in \texttt{S13} returns a value of type \texttt{S15}, which
corresponds to a selection of messages to send to \texttt{P2}. \Cref{lst:svrprotocol}
illustrates how the developer may implement these types.

We make a key design decision \textit{not} to expose communication channels in
the TypeScript session type encodings to provide \textit{static} linearity
guarantees (\cref{section:serverlinear}).
Our encoding sufficiently exposes seams for the developer to inject their
program logic, whilst the generated session API
(\cref{section:serversessionapi}) handles the sending and receiving of
messages.

\subsubsection{Session Runtime}
\label{section:serversessionapi}

The generated code for our session runtime performs communication in a protocol-conformant manner, but
does not expose these IO actions to the developer by delegating the
aforementioned responsibilities to an inner class.
The runtime executes the EFSM by keeping track of
the current state (similar to the generated code in \cite{javatypestate})
and only permitting the specified IO actions at the current state.
The runtime listens to message (receiving) events on the communication channel,
invokes the corresponding callback to obtain the value to send next, and
performs the sending.
The developer instantiates a session by constructing an instance of the
session runtime class, providing the WebSocket endpoint URL (to open the
connection) and the initial state (to execute the EFSM).

\subsubsection{Linear Channel Usage}
\label{section:serverlinear}
Developers writing their implementation using the generated APIs 
enjoy channel linearity by construction.
Our library design prevents the two conditions detailed below:

\paragraph{Repeated Usage}
We do not expose channels to the developer, which makes \textit{reusing
  channels} impossible.
For example, to send a message, the generated API only requires the payload
that needs to be sent, and the session runtime performs the send internally,
guaranteeing this action is done \textit{exactly once} by construction.

\paragraph{Unused Channels}
The initial state must be supplied to the session runtime
constructor in order to instantiate a session;
this initial state is defined
in terms of the successor states, which in turn has references to its
successors and so forth.
The developer's implementation will cover the terminal state
(if it exists), and the
session runtime guarantees this terminal state will be reached
by construction.

\subsection{The React Framework}
Our browser-side session type encodings for browser-side targets build upon the
\emph{React.js} framework, developed by Facebook \cite{React} for the
\textit{Model-View-Controller} (MVC) architecture.
React is widely used in industry to create scalable single-page TypeScript
applications, and we intend for our proposed workflow to be beneficial in an
industrial context.
We introduce the key features of the framework.

\paragraph{Components}
A component is a reusable UI element which
contains its own markup and logic.
Components implement a \texttt{render()} function which returns a React
element, the smallest building blocks of a React application, analogous to the
view function in the MVU architecture.
Components can keep \textit{state}s and the \texttt{render()} function is
invoked upon a change of state.

For example, a simple counter can be implemented as a component,
with its \texttt{count} stored as state.
When rendered, it displays a button which increments \texttt{count}
when clicked and a \texttt{div} that renders the current
\texttt{count}.
If the button is clicked, the \texttt{count} is incremented, which triggers a
re-rendering (since the state has changed), and the updated \texttt{count} is
displayed.

Components can also render other components, which gives rise
to a parent/child relationship between components.
Parents can pass data to children as \textit{props} (short for properties).
Going back to the aforementioned example, the counter component could
render a child component \texttt{<StyledDiv count=\{this.state.count\} />} in
its \texttt{render()} function, propagating the \texttt{count} from its state
to the child.
This enables reusability, and for our use case, gives control to the parent
on what data to pass to its children (e.g. pass the payload of a received
message to a child to render).

\subsection{Browser-Side API Generation}

\label{section:browser}

\begin{wrapfigure}{R}{0.5\textwidth}
  \begin{center}
    \includegraphics[width=0.5\textwidth]{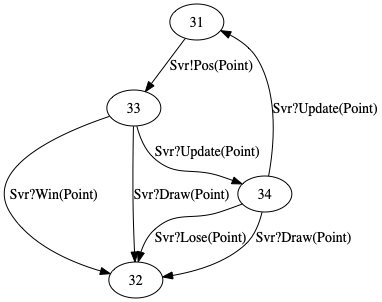}
  \end{center}

  \captionof{figure}{EFSM for \texttt{P1}.}
  \label{fig:efsmp1}
\end{wrapfigure}

We refer to the \texttt{P1} EFSM (\cref{fig:efsmp1}) as a running example in
this section.
Preserving behavioural typing and channel linearity is challenging
for browser-side applications due to EFSM transitions being triggered by user
events:
in the case of \textit{Noughts and Crosses}, once the user makes a move by
clicking on a cell on the game board, this click event must be deactivated
until the user's next turn, otherwise the user can click again and violate
channel linearity.
Our design goal is to enforce this statically through the generated APIs.

For browser-side targets, we extend the approach presented in \cite{MVU2019} on
\textit{multiple model types} motivated by the \textit{Model-View-Update} (MVU)
architecture.
An MVU application features a \textit{model} encapsulating application
state, a \textit{view function} rendering the state on the Document Object Model (DOM), and an
\textit{update function} handling \textit{messages} produced by the
rendered model to produce a new model.
The concept of model types express type dependencies between these
components: a \emph{model type} uniquely defines a \textit{view function},
set of \textit{messages} and \textit{update function} -- rather than
producing a new model, the update function defines valid transitions to
other model types.
We leverage the correspondence between model types and states in the EFSM:
each state in the EFSM is a model type, the set of messages represent
the possible (IO) actions available at that state,
and the update function defines which successor state to transition to,
given the supported IO actions at this state.

\subsubsection{Model Types in React}

\paragraph{State}
An EFSM state is encoded as an \textit{abstract} React
component.
This is an abstract class to require the developer to provide their
own view function, which translates conveniently to the \texttt{render()}
function of React components.
Our session runtime (\cref{section:clientruntime}) ``executes'' the EFSM and
renders the current state.
Upon transitioning to a successor state, the successor's view function will be
invoked, as per the semantics expressed in \cite{MVU2019}.

\paragraph{Model Transitions}
Transitions are encoded as React component props onto the encoded states by the
session runtime (\cref{section:clientruntime}).
We motivate the design choice of not exposing channel resources to provide
guarantees on channel usage.
React components in TypeScript are
\textit{generic} \cite{TypeScriptSpec}, parameterised by the permitted
types of prop and state.
The parameters allow us to leverage the TypeScript compiler to
verify that the props for model transitions stay local to the state they are
defined for.
The model transitions for EFSMs are message send and receive.

\subparagraph{Sending}
We make the assumption that message sending is triggered by
some user-driven UI event (e.g. clicking a button, pressing a key on the
keyboard) which interacts with some DOM element.
We could pass a
\texttt{send()} function as a prop to the sending state, but the developer
would be free to call the function multiple times which makes channel reuse
possible.
Instead, we pass a \textit{factory function} as a prop, which will,
given an HTML event and an event handler function, return a fresh React
component that binds the sending action on construction.
So once the bound event is triggered, our session runtime executes the event
handler function to obtain the payload to send, perform the send
\textit{exactly once} and transition to (which, in practice, means render) the
successor state.

\begin{figure}[!h]
\begin{lstlisting}[language=JavaScript, tabsize=4]
// Inside some render() function..
{board.map((row, x) => (
	row.map((col, y) => {
		const SelectPoint = this.props.Pos('click', (event: UIEvent) => {
			event.preventDefault();
			return { x: x, y: y };}
		return <SelectPoint><td>.</td></SelectPoint>;
});}
\end{lstlisting}
\captionof{lstlisting}{Model transition for message sending in
\textit{Noughts and Crosses} \texttt{P1} implementation.}
\label{lst:clientapp}
\end{figure}

We demonstrate the semantics using the \textit{Noughts and Crosses} example in
\cref{lst:clientapp}.
The session runtime passes the factory function \texttt{this.props.Pos} as a prop.
For each x-y coordinate on the game board, we
create a \texttt{SelectPoint} React component from the factory function (which
reads ``build a React component that sends the \texttt{Pos} message with x-y
coordinates as payload when the user clicks on it'') and we wrap a table cell
(the game board is rendered as an HTML table) inside the \texttt{SelectPoint}
component to bind the click event on the table cell.

\subparagraph{Receiving}
The React component for a receiving state is required to
define a handler for each supported branch. 
Upon a message receive event, the session runtime invokes the
handler of the corresponding branch with the message payload and 
renders the successor state upon completion.

\subsubsection{Session Runtime}
\label{section:clientruntime}

The session runtime can be interpreted as an abstraction on top of the React
VDOM that implements the EFSM by construction.
The session runtime itself is a React component too, named after the endpoint
role identifier:
it opens the WebSocket connection to the server, keeps track of the current
EFSM state as part of its React component state, and most importantly, renders
the React component encoding of the active EFSM state.
Channel communications are managed by the runtime, which allows it to render
the successor of a receive state upon receiving a message from the channel.
Similarly, the session runtime is responsible for passing the required props
for model transitions to EFSM state React components.
The session runtime component is rendered by the developer and requires, as
props, the \textit{endpoint URL} (so it can open the connection) and a list of
\textit{concrete state components}.

The developer writes their own implementation of each state (mainly to
customise how the state is rendered and inject business logic into state
transitions) by extending the abstract React class components.
The session runtime requires references to these concrete components in order to
render the user implementation accordingly.

\subsubsection{Affine Channel Usage}
A limitation of our browser-side session type encoding is only being able to
guarantee that channel resources are used \textit{at most once} as opposed to
\textit{exactly once}.

Communication channels are not exposed to the developer so multiple sends are
impossible.
This does not restrict the developer from binding the send action to exactly
one UI event: for \textit{Noughts and Crosses}, we bind the \texttt{Pos(Point)}
send action to each unoccupied cell on the game board, but the generated
runtime ensures that, once the cell is clicked, the send is only performed once
and the successor state is rendered on the DOM, so the channel resource used to
send becomes unavailable.

However, our approach \textit{does not} statically detect whether all
transitions in a certain state are bound to some UI event.
This means that it is possible for an implementation to \textit{not} handle
transitions to a terminal state but still type-check, so we cannot prevent
unused states. Equally, our approach does not prevent a client closing the browser, which would drop the connection.

\section{Case Study}
\label{section:example}

\begin{figure}
\begin{lstlisting}[language=JavaScript, tabsize=4]
const handleP1Move: S13 = (move: Point) => {
	board.P1(move);			// User logic
	if (board.won()) {
		return [Labels.Lose, move, [Labels.Win, move]]; (*@\label{linelose}@*)
	} else if (board.draw()) {
		return [Labels.Draw, move, [Labels.Draw, move]]; (*@\label{linedraw}@*)
	} else {
		return [Labels.Update, move, [Labels.Update, move, handleP2Move]]; 	(*@\label{lineupdate}@*)
	}
}

// Instantiate session - `handleP2Move` defined similarly as S19
new NoughtsAndCrosses.Svr(webSocketServer, handleP1Move);
\end{lstlisting}
\captionof{lstlisting}{Session runtime instantiation for \textit{Noughts and Crosses} \texttt{Svr}.}
\label{lst:svrprotocol}
\end{figure}

We apply our framework to implement a web-based implementation of the
\textit{Noughts and Crosses} running example in TypeScript;
the interested reader can find the full implementation in
\cite{NoughtsAndCrosses}.
In addition to MPST-safety, we show that our library design welcomes idiomatic
JavaScript practices in the user implementation and is interoperable with
common front- and back-end frameworks.

\paragraph{Game Server}
We set up the WebSocket server as an Express.js \cite{ExpressJS}
application on top of a Node.js \cite{NodeJS} runtime.
We define our own game logic in a \texttt{Board} class to keep track of the
game state and expose methods to query the result.
This custom logic is integrated into our \texttt{handleP1Move} and
\texttt{handleP2Move} handlers (\cref{lst:svrprotocol}), so the session runtime
can handle \texttt{Pos(Point)} messages from players and transition to the
permitted successor states (\cref{lst:game}) according
to the injected game logic: if \texttt{P1} played a winning move (\cref{linelose}),
\texttt{Svr} sends a \texttt{Lose} message to \texttt{P2} with the winning
move, and also sends a \texttt{Win} message to \texttt{P1};
if \texttt{P1}'s move resulted
in a draw (\cref{linedraw}), \texttt{Svr} sends \texttt{Draw} messages to
both \texttt{P2} and \texttt{P1}; otherwise, the game continues (\cref{lineupdate}),
so \texttt{Svr} updates both \texttt{P2} and \texttt{P1} with the latest
move and proceeds to handle \texttt{P2}'s turn.

Note that, by TypeScript's structural typing
\cite{TypeScriptSpec}, replacing
\texttt{handleP2Move} on \cref{lineupdate} with a recursive
occurrence of \texttt{handleP1Move} would be
type-correct --- this allows for better code reuse as opposed to
defining
additional abstractions to work around the limitations of nominal
typing in
\cite{Hybrid2016}.
There is also full type erasure when transpiling to JavaScript to
run the
server code, so the types defined in TypeScript will not appear in the
JavaScript after type-checking.
This means state space explosion is not a runtime consideration.

\paragraph{Game Clients}
We implement the game client for \texttt{P1} and
\texttt{P2} by extending from the generated abstract React (EFSM state)
components and registering those to the session runtime component.

For the sake of code reuse, \cite{NoughtsAndCrosses}
uses \textit{higher-order components} (HOC) to build the correct state
implementations depending on which player the user chooses to be.
We leverage the \textit{Redux} \cite{Redux} state management library to keep
track of the game state, thus showing the flexibility of our library design in
being interoperable with other libraries and idiomatic JavaScript practices.
Our approach encourages the separation of concerns between the
communication logic and program logic --- the generated session runtime
keeps track of the state of the EFSM to ensure protocol conformance by
construction, whilst \textit{Redux} solely manages our game state.

\section{Related Work}
The two main approaches for incorporating our MPST workflow into application
development are native language support for first-class linear channel
resources \cite{ATS} and code generation.
The latter closely relates to our proposal;
we highlight two areas of existing work under this approach that motivate our
design choice.

\paragraph{Endpoint API Generation}
Neykova and Yoshida targeted Python applications and the generation of runtime
monitors \cite{Python2017} to dynamically verify communication patterns.
Whilst the same approach could be applied to JavaScript, we can provide more
static guarantees with TypeScript's gradual typing system.
Scribble-Java \cite{Hybrid2016} proposed to encode the EFSM
states and transitions as classes and instance methods respectively, with
behavioural typing achieved statically by the type system and channel linearity
guarantees achieved dynamically since channels are exposed and
aliasing is not monitored.
Scribble-Java can generate callback-style APIs similar to the approach we
present, but this approach is arguably less idiomatic for Java developers.

\paragraph{Session Types in Web Development}
King et al.\ \cite{PureScript2019} targeted web development in PureScript using the
\textit{Concur UI} framework and proposed a type-level encoding of EFSMs as
multi-parameter type classes.
However, it presents a trade-off between achieving static linearity guarantees
from the type-level EFSM encoding under the expressive type system and
providing an intuitive development experience to developers, especially given
the prevalence of JavaScript and TypeScript applications in industry.
Fowler \cite{MVU2019} focused on applying binary session types in front-end web
development and presented approaches that tackle the challenge of guaranteeing
linearity in the event-driven environment, whereas our work is applicable to
multiparty scenarios.

Our work applies the aforementioned approaches in a \textit{multiparty} context
using industrial tools and practices to ultimately encourage MPST-safe web
application development workflows in industry.


\section{Conclusion and Future Work}
We have presented an MPST-based framework for developing full-stack interactive
TypeScript applications with WebSocket communications.
The implementation conforms to a specified
protocol, statically providing linear channel usage guarantees and affine
channel usage guarantees for back-end and front-end targets respectively.

Future work includes incorporating \textit{explicit connection actions}
introduced in \cite{ExplicitConnections} in our API generation to better model
real-world communication protocols that may feature in interactive web
applications.
Server-side implementations may perform asynchronous operations on the
received messages, so supporting asynchronous values (such as JavaScript
\textit{Promises} \cite{promise}) in IO actions would be a welcome addition.
Whilst our approach supports multiparty sessions, the nature of
WebSockets require some server-based role in the communication protocol and
clients to interact via the server.
Extending support to WebRTC \cite{WebRTC} would
cater for peer-to-peer communication between browsers, which further opens up
possibilities for communication protocols supported by our approach.



\section*{Acknowledgements}
We thank the anonymous reviewers for their 
feedback.\\
This work was supported in part by EPSRC projects EP/K011715/1, EP/K034413/1, EP/L00058X/1, EP/N027833/1, EP/N028201/1, and EP/T006544/1.
\nocite{*}
\bibliographystyle{eptcs}
\bibliography{generic}
\end{document}